\documentclass{aa}
\usepackage{graphics}

\begin{document}

\title{H$_2$, HD, and CO at the edge of 30~Dor in the LMC: \\ the line of sight to Sk\,-69\,246\thanks{Based on observations made with the NASA-CNES-CSA Far Ultraviolet 
Spectroscopic Explorer, available in the public archive. FUSE is operated for NASA by the Johns Hopkins 
University under NASA contract NAS5-32985.}}

\author{H. Bluhm and K.S. de Boer}

\institute{Sternwarte, Universit\"at Bonn, Auf dem H\"ugel 71, 
D-53121 Bonn, Germany}

\date{Received <date> / Accepted <date>}

%\thesaurus{08(09.01.1;09.13.2;10.19.3;13.21.3)}

\offprints{H. Bluhm
\email{hbluhm@astro.uni-bonn.de}
}

\authorrunning{H. Bluhm \& K.S. de Boer}
\titlerunning{H$_2$, HD, and CO at the edge of 30~Dor}

\abstract{
{\sc Fuse} and {\sc IUE} observations of \object{Sk\,-69\,246}, a WN\,6 star to the north of \object{30~Dor}, reveal the existence of \object{LMC} interstellar gas in at least 3 velocity components. In one of these components {\sc Fuse} spectra show strong absorption by molecular hydrogen with a column density of $\log N$(H$_2$)$=19.6$.
 We investigate the physical conditions in this cloud, which is probably located near the \ion{H}{ii} region surrounding \object{Sk\,-69\,246}.
 HD and CO with column densities of $\approx 13.9$ and $13.0$, respectively, are detected.
 The $N$(CO)-to-$N$(H$_2$) ratio is consistent with values found in the Galaxy.
 From the population of the rotational levels of H$_2$ we derive a gas temperature of $72$~K and a radiation field $U\approx3\cdot10^{-15}$~erg\,s$^{-1}$\,cm$^{-3}$ at 1000~{\AA}.
 The fractional abundance $f=2N({\rm H}_2)/[N({\rm \ion{H}{i}})+2N({\rm H}_2)]\approx0.07$ is rather high for an $E(B-V)\la0.1$.
}

\maketitle

\section{Introduction}
 The 30 Dor complex of young stars and emission nebulosity in the \object{LMC} has been studied in almost every wavelength domain. 
 The detected X-ray flux indicated the presence of very hot gas (Wang \cite{wang}). 
 The UV, visual, and near IR part of the spectrum have been used to study the stars (e.g. Walborn et al. \cite{walborn}, Selman et al. \cite{selman}), the interstellar absorption lines due to cool gas (e.g. de Boer et al. \cite{deboer80}, Vladilo et al. \cite{vladilo}) as well as ionized gas (de Boer \& Savage \cite{deboer80b}). 
 In the radio domain the cool component of the gas was studied with the 21 cm emission of H\,{\sc i}
(e.g. Rohlfs et al. \cite{rohlfs} and Luks \& Rohlfs \cite{luks}) and molecular emission lines (see e.g. Heikkil{\"a} et al. \cite{heikkila}).

 Since the stars in the 30 Dor complex are young, while the emission by the gas indicates a well ionized volume with warm or hot gas, it is of interest to investigate how much of the original star-forming cloud is left in that region. 
 Measurements in CO (Mizuno et al. \cite{mizuno}) indicate that only limited amounts of that molecule are present within a radius of $30\arcmin$ from 
\object{R~136}, the brightest central object. 
 The structure of the gas in the \object{30~Dor} region is rather complex. 
 Several velocity components are known to exist, as followed both from investigations in the visual (Vladilo et al. \cite{vladilo}) and in the ultraviolet (de Boer et al. \cite{deboer80}). 
 
 Investigations of molecular hydrogen, H$_2$, have become possible since 1996. 
 In that year, the {\sc Orfeus} space shuttle based far-UV spectrograph collected for the first time spectra of stars in the Magellanic Clouds in the wavelength range of 900 - 1200~{\AA}, the wavelength region in which H$_2$ absorption lines are present. 
 Since 1999 also the {\sc Fuse} satellite has been used to explore this part of the spectrum. 
 The first detection of interstellar H$_2$ in absorption in \object{LMC} gas (de Boer et al. \cite{dBrichter}) was followed by a first overall appraisal of H$_2$ in \object{LMC} gas (Richter \cite{richter2000}). 

 Here we wish to explore the nature of the molecular gas, in particular that of H$_2$, at the edge of the 30 Dor complex. 
 We have selected the line of sight to \object{Sk\,-69\,246} (= \object{HD~38282} = \object{R~144}), a line of sight for which the interstellar gas was investigated by de Boer et al. (\cite{deboer80}) using {\sc IUE} data. 
 \object{Sk\,-69\,246} is a WN6 star 4{\arcmin} north of \object{R~136}, the star cluster in the centre of \object{30~Dor}. Crowther \& Smith (\cite{crowther}) derive $M_V=-7.75$~mag and $E(B-V)=0.1$~mag, including a galactic reddening of $0.05$~mag.
 From the {\sc Fuse} spectrum we derive a luminosity $\frac{\partial^2 E}{\partial t \partial \lambda}\approx 2.4\cdot 10^{36}$~erg\,s$^{-1}$\,{\AA}$^{-1}$ at $\lambda\approx 1000$~{\AA}.

\section{Data}
 {\sc Fuse} is equipped with four coaligned telescopes and Rowland spectrographs.
 The detectors are two microchannel plates.
 For a description of the instrument and its on-orbit performance see Moos et al. (\cite{moos}) and Sahnow et al. (\cite{sahnow}). 
 We analysed {\sc Fuse} data of \object{Sk\,-69\,246} observed on Dec. 16, 1999 (with the identifier P1031802000) calibrated with the Calfuse 1.5.3 reduction pipeline.
 Four single exposures are available, with exposure times of 3453, 7393, 6414, and 4817 s.
 Adding these exposures (each channel seperately) improved the data quality, but some noise patterns remain essentially unchanged compared to the single exposures.

 The spectral resolution of the data is uncertain. 
 Our fits worked best with an instrumental FWHM of about 25 - 30~km\,s$^{-1}$, a value significantly higher than the aim of the {\sc Fuse} project of $10$~km\,s$^{-1}$.
 The actual resolution might be better than stated above if the line profiles are broadened by an unrecognized velocity structure.
 An obvious problem of the handling of the {\sc Fuse} spectra is the alignment of the 8 different detector channels which affects the measurement of radial velocities.
 While zero points and velocity differences are largely consistent within each channel (except for some detector edge regions), the zero points differ hugely (up to $\approx 80$~km\,s$^{-1}$) between the channels.
 We thus have to regard radial velocities from the {\sc Fuse} spectrum as arbitrary and rely only on velocity differences. 
 A rough calibration of the velocity zero point can be made by comparison to \ion{H}{i} emission line data. 
  
 The quality of the background subtraction and the countrates (in spite of a bin size of only $\approx 2$~km\,s$^{-1}$) are excellent.
 They surpass the {\sc Orfeus} echelle spectra of \object{LMC} stars, which were the only FUV intermediate resolution data available before the {\sc Fuse} mission.   

 In addition to the {\sc Fuse} data we used {\sc IUE} spectra (SWP 02798, LWR 03766) from the public archive. 
 Though the {\sc IUE} data available for \object{Sk\,-69\,246} are only of mediocre quality, with all known problems like low signal-to-noise ratio, sometimes dubious background subtraction or uncertain radial velocity zero point, they are a valuable complement to the {\sc Fuse} data.

\section{Velocity structure and column densities}
\subsection{Metals}

\begin{figure}
\resizebox{\hsize}{!}{\includegraphics{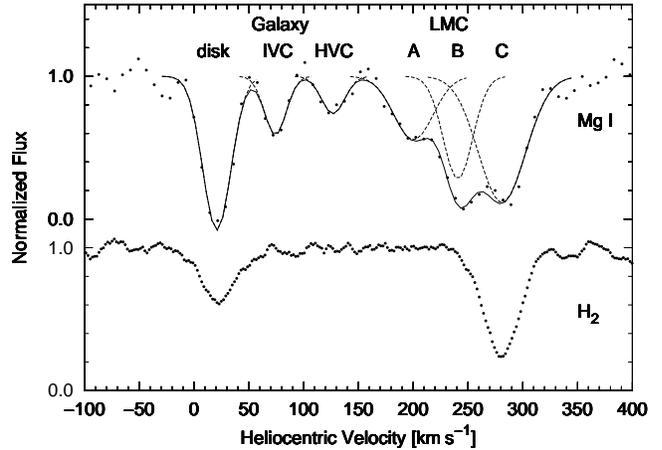}}
\hfill
\caption[]{The velocity structure in transitions of \ion{Mg}{i} (2852~{\AA}, {\sc IUE}) and H$_2$ (Ly P(3)(15-0), 944~{\AA}, {\sc Fuse}). The components fitted to the \ion{Mg}{i} absorption are at $v_{\rm hel}\approx20$~km\,s$^{-1}$ (disk gas), at $70$ and $125$~km\,s$^{-1}$ (IVC and HVC gas) and the components A ($200$~km\,s$^{-1}$), B ($240$~km\,s$^{-1}$), and C ($280$~km\,s$^{-1}$) discussed in this paper. The full line gives an overall fit. 
In the H$_2$ transition, absorption is seen at galactic disk and component C velocity 
}
\label{mgh2}
\end{figure}

\begin{figure}
\resizebox{\hsize}{!}{\includegraphics{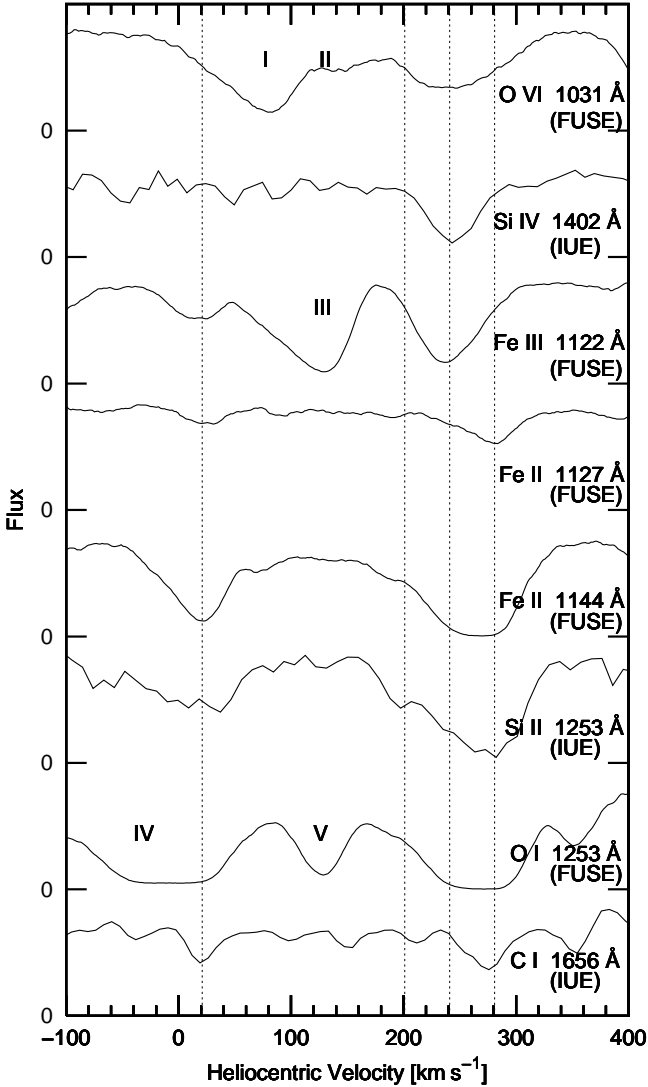}}
\hfill
\caption[]{The velocity structure in transitions of different metals. The dotted lines mark the positions of galactic absorption at $v_{\rm hel}\approx20$~km\,s$^{-1}$ and of our components A,B, and C at $v_{\rm hel}\approx200$, $240$, and $280$~km\,s$^{-1}$. Absorptions by other transitions blending into the velocity range are: {\sc I}) H$_2$ Ly P(3)(6-0) [LMC]; {\sc II}) H$_2$ Ly R(4)(6-0) [gal.]; {\sc III}) \ion{Fe}{ii} 1121~{\AA} [LMC]; {\sc IV}) H$_2$ Ly P(1)(5-0) [LMC]; {\sc V}) H$_2$ Ly P(2)(5-0) [gal.]. Further absorption profiles in the {\sc IUE} spectrum are given in Savage \& de Boer (\cite{savdB}) 
}
\label{velstruc}
\end{figure}

\begin{figure}
\resizebox{\hsize}{!}{\includegraphics{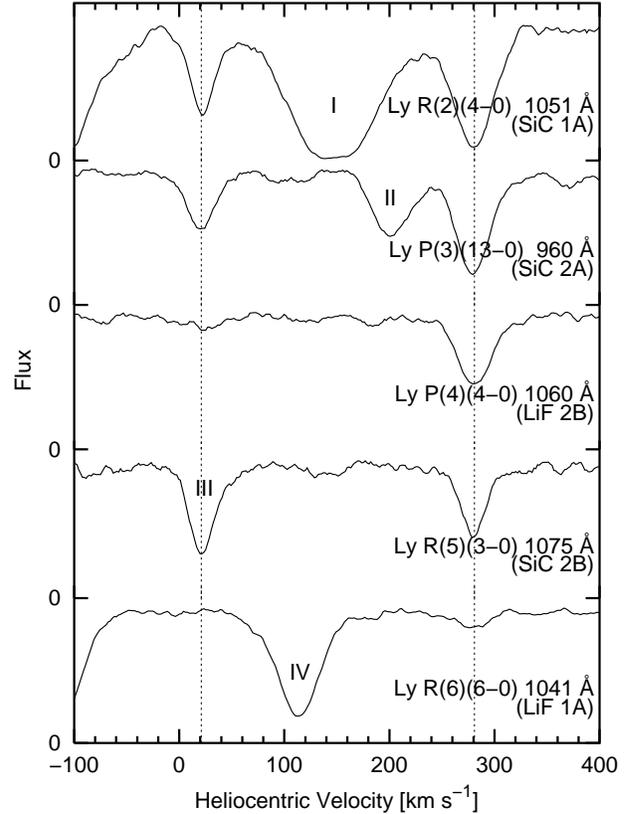}}
\hfill
\caption{H$_2$ absorption profiles of transitions of the rotational states $J=2$ to $J=6$. The labels give the name of a transition, its approximate wavelength, and the detector segment. Roman numbers mark \object{LMC} absorptions belonging to other transitions. {\sc I}) Ly~P(1)(4-0); {\sc II}) Ly~P(5)(14-0); {\sc III}) Ly~P(4)(3-0); {\sc IV}) Ly~R(3)(5-0)} 
\label{h2vel}
\end{figure}

\begin{table}
\caption[]{\object{LMC} column densities $\log N$ [cm$^{-2}$] }
\label{colden}
\setlength{\tabcolsep}{0.7mm}
\begin{flushleft}
\begin{tabular}{lllll}
\hline\noalign{\smallskip}
         & Cloud A & Cloud B & Cloud C & total \\
$v_{\rm hel}$ [km\,s$^{-1}$] & $+200$ & $+240$ & $+280$  &  \\
\noalign{\smallskip}\hline\noalign{\smallskip}
\ion{S}{ii}  & $14.75\pm0.2$ & $15.0\pm0.2$ & $15.6\pm0.3$  & $15.75\pm0.2$ \\
\ion{Fe}{ii} & $13.8\pm0.1$  & $14.7\pm0.1$ & $15.2\pm0.05$ & $15.3\pm0.05$ \\ 
\ion{O}{i}   & & & & $16.8^{+0.6}_{-0.4}$ \\
%\ion{H}{i}   & & & & $21.4\pm0.2$ \\
H$_2$        & -- & -- & $19.63^{+0.04}_{-0.08}$ & \\
HD ($J=0$)   & -- & -- & $13.9\pm0.3$ & \\
HD ($J=1$)   & -- & -- & $<13.5$ & \\
CO           & -- & -- & $13.0\pm0.4$ \\

\noalign{\smallskip}\hline\noalign{\smallskip}
\end{tabular}

\end{flushleft}
\end{table}

 Absorption by neutral and low ionized species shows the velocity structure common to \object{LMC} sight-lines: galactic absorption by disk gas near 20~km\,s$^{-1}$ (heliocentric), by intermediate velocity gas at $\approx +80$~km\,s$^{-1}$ and high velocity gas at $\approx +130$~km\,s$^{-1}$, in addition to \object{LMC} absorption between $\approx +200$ and $+300$~km\,s$^{-1}$.
 For the classification of the velocity components as galactic or magellanic see Wayte (\cite{wayte}).
 Relatively weak \object{LMC} absorption lines of \ion{S}{ii}, \ion{Fe}{ii}, or \ion{Mg}{i} show a three component structure at heliocentric velocities of $\approx 200$, $240$, and $280$~km\,s$^{-1}$, here denoted as component A, B, and C.
 In strong transitions these components are not resolved.
 Fig.~\ref{mgh2} shows the basic structure in the spectra, including spectral fits. 
 Further examples of absorption lines are given in Fig.~\ref{velstruc}.    
 Of course, a resolution of $20$ to $30$~km\,s$^{-1}$ is low enough to hide some substructure.
 Indeed, optical spectra of \ion{Na}{i} and \ion{Ca}{ii} at a FWHM of $\approx 3$ - $7$~km\,s$^{-1}$ can be fitted with two components seperated by about $8$~km\,s$^{-1}$ (see Vladilo et al. \cite{vladilo}) which are hidden in our component C. 
 The radial velocities given above are based on the identification of the \ion{Na}{i} velocity with that of H$_2$.
 \object{LMC} column densities for a few metal species are listed in Table \ref{colden}.
 In some cases line blending did not allow the determination for each velocity component seperately.

\subsection{Molecular hydrogen}

\begin{table}
\caption[]{H$_2$ column densities $N$ [cm$^{-2}$] in the \object{LMC} component C at $\approx280$~km\,s$^{-1}$. 
 Given are the values for different rotational states with quantum number $J$.
 Apart from the uncertainties in the equivalent widths or the scatter in the profile fits, the errors take the possibility of a deviating curve of growth into account}
\label{H2cd}
\begin{flushleft}
\begin{tabular}{lllll}
\hline\noalign{\smallskip}
$J$ & 0 & 1 & 2 & 3 \\
$\log N$ & $19.36^{+0.03}_{-0.07}$ & $19.28^{+0.08}_{-0.08}$ & $17.5^{+0.4}_{-0.5}$ & $17.0^{+0.6}_{-0.5}$ \\
\noalign{\smallskip}\hline\noalign{\smallskip}
$J$ & 4 & 5 & 6 & 7 \\
$\log N$ & $15.7^{+0.2}_{-0.3}$ & $15.2^{+0.1}_{-0.3}$ & $14.15^{+0.2}_{-0.2}$ & $<13.5$ \\
\noalign{\smallskip}\hline\noalign{\smallskip}
\end{tabular}
\end{flushleft}
\end{table}

 Molecular hydrogen is solely detected in the galactic disk component and \object{LMC} component C.
 For the \object{LMC}, H$_2$ lines belonging to rotational states from $J = 0$ to $J = 6$ can be measured.
 Because radiative transitions between the rotational levels within the electronic and vibrational ground state are forbidden, the population of these levels is governed by collisional excitation or de-excitation and UV-pumping (see Sect. 4.2.)
 In principle, the two H$_2$ absorption bands accessible in the UV, Lyman and Werner, provide a  wealth of transitions for each rotational level, but for high column densities and complex velocity structures, this blessing turns into a curse.
 Especially at wavelengths below $1000$~\AA\ the number of H$_2$ lines is so large that many of them can hardly be used for analysis because of severe blending.

\begin{figure}
\resizebox{\hsize}{!}{\includegraphics{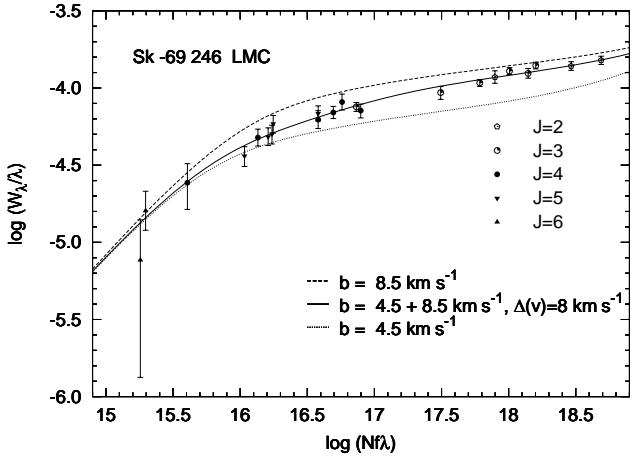}}
\hfill
\caption[]{Optical depth relation for H$_2$ in component C. The dashed curves are maxwellian single-cloud curves of growth with $b$-values of 5.5 and 8.5~km\,s$^{-1}$, the solid curve is the optical depth relation as given by the model described in Sect. 3.2 
}
\label{cog}
\end{figure}

 For $J=2$ to $J=6$ equivalent widths were measured. 
 Their errors were calculated from photon statistics and an estimate of the continuum uncertainty.
 Column densities were determined with the curve of growth method.
 It turns out that a maxwellian single-cloud curve of growth does not fit the \object{LMC} H$_2$ data.
 The optical depth function is changed either by an unresolved velocity structure or by a non-maxwellian velocity distribution within a velocity component.
  Note that the curve is not the result of some kind of maximum likelihood calculation.
 For each rotational level the measured equivalent widths form a set of data points with unknown position on the $\log(Nf\lambda)$ axis.
 These sets have to be shifted along this axis in order to form a monotonously rising curve from which the column densities follow.
 For constructing a model optical-depth-relation we chose a two-component-model as simplest approach. 
 We adopted a radial velocity difference of 8~km\,s$^{-1}$ as fitted by Vladilo et al. (\cite{vladilo}) to \ion{Na}{i} lines and computed optical-depth-relations for different column densities and $b$-values. 
 The model that we chose finally has $4.5$ and $8.5$~km\,s$^{-1}$ as Doppler parameters, with the latter component having a column density lower by one order of magnitude.  
  The column densities given in Table \ref{H2cd} are total \object{LMC} values with errors taking into account the errors in the equivalent widths and an estimate of systematic uncertainties arising from the optical depth relation.

\begin{figure}
\resizebox{\hsize}{!}{\includegraphics{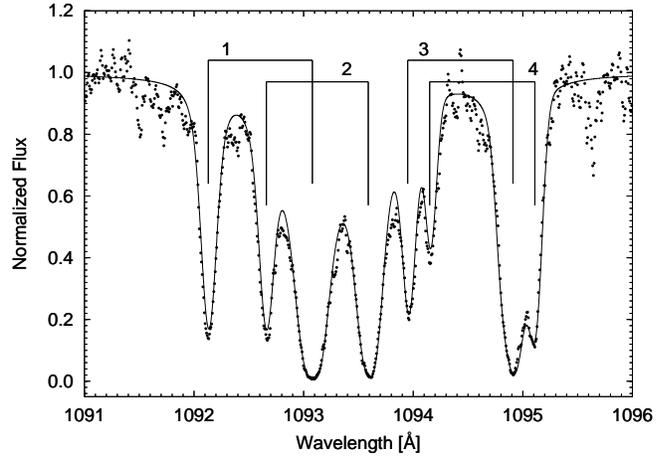}}
\hfill
\caption[]{For some H$_2$ absorption lines belonging to the 1-0 band, computed profiles (solid line) are overlaid to the {\sc Fuse} spectrum (data points). The brackets mark galactic and \object{LMC} absorptions belonging to the same transition. {\bf 1}: Lyman R(0), {\bf 2}: Lyman R(1), {\bf 3}: Lyman P(1), {\bf 4}: Lyman R(2)}
\label{1-0}
\end{figure}

 The absorption lines belonging to the $J=0$ and $J=1$ levels are strong enough to allow Voigt profile fitting.
 In these lines the component with the lower $b$-value and the higher column density is clearly dominating the profile.
 When deriving the total column density, one- or two-component fits make for only a minimal difference. 
 We used lines from 5 absorption bands (0-0 to 4-0) for $J=0$ and $J=1$.
 In the fits the dispersion in the velocity difference between galactic and \object{LMC} components is $2.9$~km\,s$^{-1}$, the 8~km\,s$^{-1}$ velocity difference between the two \object{LMC} components has been held.  
 Fig. \ref{1-0} shows how well the measured profiles can be fitted.

 The total amount of H$_2$ of all levels given in Table \ref{H2cd} is $\log N$(H$_2$)$=19.63^{+0.04}_{-0.08}$.

\subsection{Other molecules}
 Component C contains enough H$_2$ that the search for other molecules might be rewarding.
 Two sorts of relatively abundant molecules have absorption lines in the {\sc Fuse} wavelength range: carbon monoxide and deuterated hydrogen.
 Indeed both kinds of molecules are detected, for the first time in absorption in the \object{LMC}.

{\sl HD:}\\ 
 The HD lines are very weak, with equivalent widths of $W_{\lambda}\approx10$~m{\AA}, not much larger than what might be fixed pattern features.
 To gain some confidence in the detection, we compared the \object{Sk\,-69\,246} spectrum to that of \object{Sk\,-65\,22}, which is also public in the {\sc Fuse} archive and has also been reduced with Calfuse 1.5.3.
 Towards \object{Sk\,-65\,22}, H$_2$ is seen at \object{LMC} velocity, but the column density is $\log N$(H$_2$)$\approx14.9$, far lower than towards \object{Sk\,-69\,246}, thus neither HD nor CO is expected in any detectable amount. 
 If for example a feature in the \object{Sk\,-69\,246} spectrum appears at the same place (i.e. pixels) and with similar shape in the \object{Sk\,-65\,22} spectrum, it must be suspected to be an artifact.
 Such features are absent.
 
\begin{figure}
\resizebox{\hsize}{!}{\includegraphics{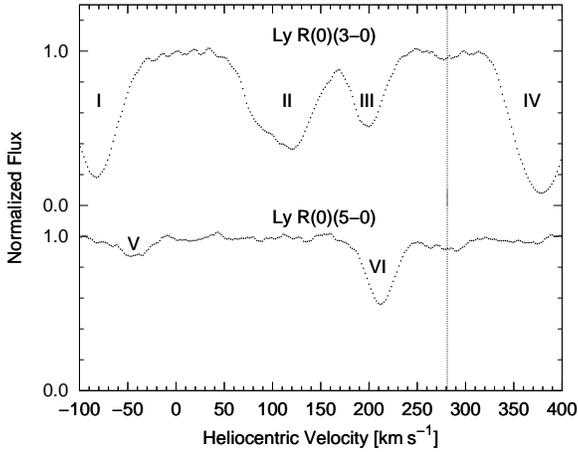}}
\hfill
\caption{HD absorption profiles belonging to transitions from the rotational ground state, Ly R(0)(3-0) at 1066.271~{\AA} and Ly R(0)(5-0) at 1042.847~{\AA}.
The radial velocity of component C is marked with the vertical dashed line. Other absorptions in the plotted spectral sections are by: {\sc I}) H$_2$ Ly R(2)(3-0) [LMC]; {\sc II}) H$_2$ Ly P(5)(4-0) [LMC] + \ion{Ar}{i} [gal.]; {\sc III}) Ly P(2)(3-0) [gal.]; {\sc IV}) H$_2$ Ly R(3)(3-0) [gal.] + \ion{Ar}{i} [LMC]; {\sc V}) H$_2$ Ly R(6)(6-0) [LMC]; {\sc VI}) H$_2$ Ly P(3)(5-0) [gal.]
}
\label{hd}
\end{figure}

 Two lines belonging to the $J=0$ level of HD are detected, Ly R(0)(3-0) and Ly R(0)(5-0) (see Fig. \ref{hd}).
 Other lines are too weak or blended. 
 For $J=1$ only upper limits can be given.
 We measure $\log N(J=0)=13.9\pm0.3$ and $\log$(HD/H$_2$)$\approx-5.7$.
 This HD-to-H$_2$ ratio is larger than that measured towards $\zeta$~Oph ($-6.3$, see Wright \& Morton \cite{wrightmort}).
 A determination of the deuterium abundance would, however, require knowledge of the exact degree of ionization in cloud C (see Black \& Dalgarno \cite{blackdal73}).  

{\sl CO:}\\
 CO has several absorption bands in the {\sc Fuse} wavelength domain.
 These bands contain transitions from different rotational levels.
 It is useful that the sums of $f$-values of transitions from these levels are constant within a band.
 So the total equivalent width of a band is independent of the population of the rotational levels.

\begin{figure}
\resizebox{\hsize}{!}{\includegraphics{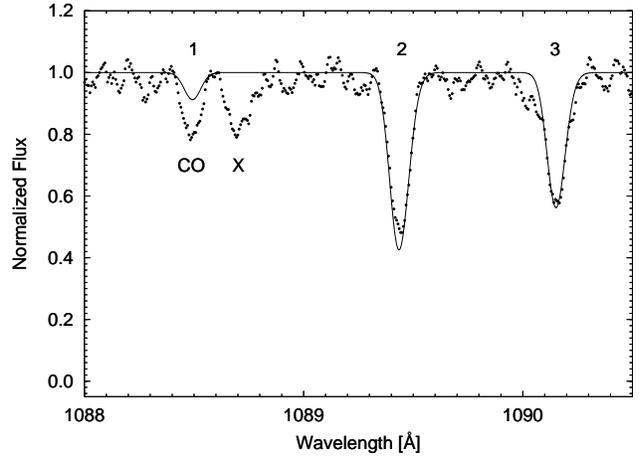}}
\hfill
\caption[]
{The section of the SiC 2b spectrum, where absorption by CO is detected. Profiles of H$_2$ absorption (solid line) modeled using the parameters derived from other transitions are overlaid to the {\sc Fuse} spectrum (data points).
These lines are:
1) The Galactic H$_2$ Ly P(4)(2-0) line. It is by far too weak to fill in the measured absorption profile. The excess absorption must be by the strongest transitions of the CO CX (1-0) band at the radial velocity of component C.
2) \object{LMC} H$_2$ Ly P(4)(2-0), 3) \object{LMC} H$_2$ Ly P(5)(2-0), the corresponding Galactic line is too weak for detection. X marks the unidentified absorption feature (see Sect.~3.3)
} 
\label{CO}
\end{figure}

 The strongest bands are the E-X (0-0) band between 1075.9 and 1076.2 {\AA} and the C-X (0-0) band between 1087.8 and 1088.0 {\AA}.
 The \object{LMC} E-X (0-0) band absorption is blended with the strong galactic H$_2$ Ly R(0)(2-0) line at 1077.138~{\AA} and thus not measurable, while the strongest transitions of the  C-X (0-0) band lie at the wavelength of the galactic H$_2$ Ly P(4)(2-0) line.  
 All other bands are by far weaker and also often blended.
 
 From equivalent width measurements of galactic H$_2$ absorption lines we derive a $b$-value of about $3$~km\,s$^{-1}$ and for $J=4$ a column density of $\log N = 14.3\pm{0.15}$.
 A comparison between a modeled galactic Ly P(4)(2-0) line and the absorption feature in the spectrum shows that additional absorption is present in the profile (see Fig. \ref{CO}). 
 From the fact that the absorption profile is narrow we conclude that almost the entire CO equivalent width is due to the C-X R(0)(0-0) line at 1087.868~{\AA} and perhaps the R(1)(0-0) and P(1)(0-0) lines at 1087.821 and 1087.959~{\AA}.
 
 In principle, this part of the spectrum is measured on three detector segments, SiC 1A, SiC 2B, and LiF 2A.
 In SiC 1A and LiF 2A the detector edge is very near the absorption line, leading to obvious problems with the wavelength calibration and fixed pattern structures, as the comparison with the \object{Sk\,-65\,22} spectrum reveals. 
 The SiC 2B measurement seems to be the most reliable although the countrate is significantly lower than that of the LiF 2A segment.

 As CO equivalent width we take the difference of the total and the H$_2$ equivalent width and derive $W_{\lambda} = 10.5\pm{7.2}$~m{\AA}.
 To calculate the error we added the following contributions quadratically:
1) the uncertainty of the H$_2$ equivalent width 2) possible blending of the intrinsic line profiles 3) photon statistics 4) the equivalent width of a "typical" probably non-photonic noise peak.
 The resulting CO column density is $\log N$(CO)$= 13.0\pm0.4$.

 Close to the CO-line a line of unknown origin is detected.
 Comparison to the other detector elements and the \object{Sk\,-65\,22} spectrum suggests that this feature is real.
 Morton (\cite{morton78}) reports the detection of an unidentified line towards $\zeta$ Oph and $\gamma$ Ara at a rest wavelength of about 1088.05~{\AA}.
 This line, Doppler shifted with the velocity of component C, could explain the feature.

\subsection{Neutral hydrogen}
 Unfortunately it seems impossible to determine a reliable column density of neutral hydrogen.
 In all lines of the Lyman series, galactic and \object{LMC} absorption is blended and
the fully damped Lyman $\alpha$ and $\beta$ lines are deteriorated by stellar lines.
 For the higher Lyman lines, fits would have to take the $b$-value into account, a procedure which is doomed to fail because of other lines, mostly of H$_2$, blending into the important flanks of the profiles.
 Even though there is the prominent \ion{N}{v} P-Cygni profile at its right wing, we tried to fit the Lyman $\alpha$ line.
 We used the value of  $N$(\ion{H}{i})$_{\rm Gal.}=0.55\cdot10^{21}$~cm$^{-1}$ for the galactic hydrogen column density given by Savage \& de Boer (\cite{savdB81}) and fitted the \object{LMC} component.      
 The result is, depending on the choice of the continuum, $N$(\ion{H}{i})$_{\rm LMC}=(2.0\pm0.5)\cdot10^{21}$~cm$^{-1}$ for the sum of all three \object{LMC} components.
 This value is roughly consistent with that derived by de Boer et al. (\cite{deboer80}) for  $N$(\ion{H}{i})$_{\rm Gal.+LMC}$=($1.9\pm0.7\cdot10^{21}$)~cm$^{-2}$ from a Ly $\alpha$ equivalent width measurement utilizing the short-wavelength half of the line.
 The \ion{H}{i} contour map by Luks \& Rohlfs (\cite{luks}) based on Parkes 21~cm emission line data shows a total \object{LMC} column density of about $3\cdot10^{21}$~cm$^{-2}$ in the direction of \object{Sk\,-69\,246}.
 Obviously a significant amount of neutral gas is located behind the star.

 Perhaps the most reliable estimates of H column densities are based on the \ion{S}{ii} column densities assuming an \object{LMC} abundance of $\log({\rm S}/{\rm H})=-5.3$ (see Russell \& Dopita \cite{russell}) and no deposition onto dust, which is justified for low $E(B-V)$. 
 Using our metal column densities we obtain $N({\rm H})=(1.1^{+0.7}_{-0.4})\cdot10^{21}$~cm$^{-2}$ for all three \object{LMC} components and $N({\rm H})=(0.8^{+0.8}_{-0.4})\cdot10^{21}$~cm$^{-2}$ for component C. 
 Note that these column densities might include ionized hydrogen because \ion{S}{ii} has a higher ionization potential than \ion{H}{i}.

\subsection{High ions}
 The \ion{O}{vi} absorption line at 1031~{\AA} is detected in the {\sc Fuse} spectrum. 
 Because of a nearby stellar emission or P-cygni feature the continuum is rather uncertain, so that shape, exact position, and depth of the line profile are difficult to assess.
 The central velocity is similar to that of component B.
 The second line of that \ion{O}{vi} doublet at 1037~{\AA} is blended with H$_2$ absorption.
 From the one \ion{O}{vi} line we estimate $\log N$(\ion{O}{vi}) $\ga13.7$. 
 In the {\sc IUE} spectrum rather sharp interstellar \ion{Si}{iv} and \ion{C}{iv} absorption lines
 are found near the radial velocity of component B, but are absent in components A and C.
 Absorption by \ion{Fe}{iii} is detected in the {\sc Fuse} spectrum, predominantly in component B (see Fig. \ref{velstruc}).
 The profile is similar to that of the \ion{Si}{iv} lines, but a bit widened to higher velocities.
 It is the only well accessible \ion{Fe}{iii} absorption line, so we can give a lower limit for the column density, $\log N$(\ion{Fe}{iii})$>14.3$ in component B.
 Details of \ion{Si}{iv} and \ion{C}{iv} lines have been presented by de Boer et al. (\cite{deboer80}).

\section{What's going on out there?}
\subsection{In general}
 The lines of highly ionized species in component B originate in the hot ionized  gas probably surrounding \object{Sk\,-69\,246}.
 Fig. \ref{velstruc} shows that in \ion{Fe}{iii} absorptions component B is the strongest, while component C is stronger in \ion{Fe}{ii}.   
 That component B is seen in \ion{Mg}{i} does not necessarily imply that there is cool gas at this velocity.
 Due to dielectronic recombination the recombination coefficient of \ion{Mg}{i} at $10^4$~K is larger than at $10^2$~K leading to an enhanced presence at $\approx10^4$~K. 
 
 Cloud A is probably situated in the far foreground of clouds B and C.
 If it was nearer to \object{Sk\,-69\,246} than the other clouds, absorption by higher ionized species should be stronger.

 The radial velocity of \object{Sk\,-69\,246} derived by Crowther \& Smith (\cite{crowther}) is given as $240$~km\,s$^{-1}$, but with a large error, as indicated by their claim that this is in good agreement with the $267$~km\,s$^{-1}$ from Moffat (\cite{moffat}), in both cases however without indicating wether it is heliocentric or LSR.
 Therefore it is uncertain whether the stellar radial velocity resembles more that of component B or C.

 Cloud C might be a remnant of the molecular cloud which gave birth to \object{Sk\,-69\,246} or the young stellar population inside \object{30~Dor} in general.
 The kinetic temperature of the molecular gas is $72\pm{7}$~K (see Sect. 4.2), which is clearly warmer than in a dark molecular cloud but still quite cool for gas exposed to a strong radiation field as it is expected in this region of the \object{LMC}.

\subsection{In particular}

\begin{figure}
\resizebox{\hsize}{!}{\includegraphics{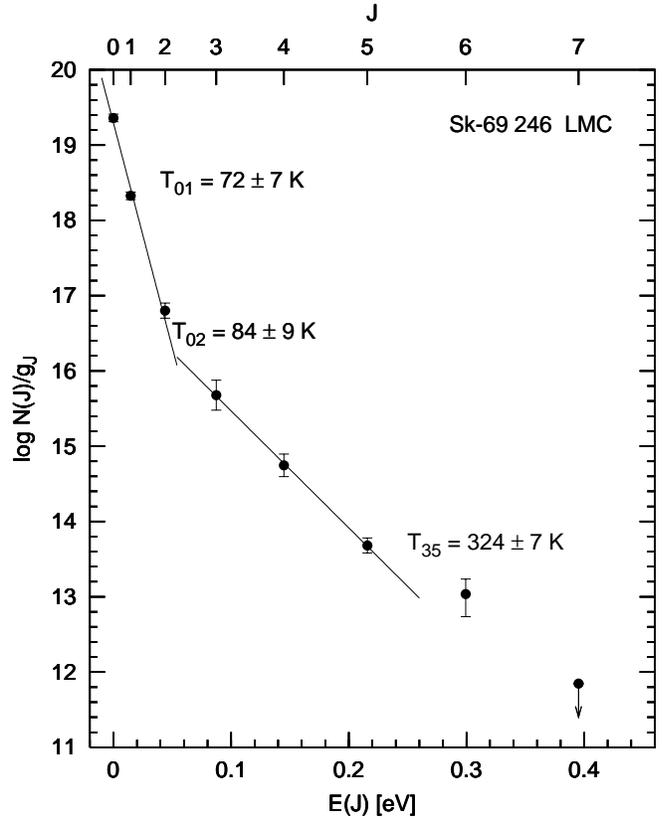}}
\hfill
\caption[]{Excitation plots for H$_2$ in component C. The lines are fits to the $J=0, 1, 2$ and $J=3, 4, 5$ data assuming a Boltzmann distribution of the level populations }
\label{temperature}
\end{figure}

 Fig. \ref{temperature} shows the H$_2$ column densities of different rotational levels divided by their statistical weights versus the level energies.
 Assuming a Boltzmann distribution, the excitation temperature is defined as 
\begin{equation}
T_{mn}=\frac{E_n-E_m}{2.303 k}\frac{1}{\log(N_m/g_m)-\log(N_n/g_n)}.
\end{equation} 
 It is a common phenomenon that at least two Boltzmann temperatures are needed to fit the distribution.
 The levels with $J=0,1,2$ are predominantly populated by collisional processes,
 the derived $T_{01}=72$~K is typical for normal neutral gas (see e.g. Savage et al. \cite{savage77} and Bluhm et al. \cite{bluhm}).
 The higher excitation temperatures of higher rotational levels are ususally explained with UV-pumping (see e.g. Spitzer \& Zweibel \cite{spitzweib}).
 After the absorption of an UV photon, an H$_2$ molecule is in an excited electronic state, usually in addition to rotational and vibrational excitation. 
 It then falls back to the electronic ground state, but is in general still vibrationally and rotationally excited.
 A cascade of so-called ro-vibrational transitions follows until the molecule reaches a rotational level in the vibrational ground state.
 Another process for populating high $J$-levels is probably molecule formation.
 For details see e.g. Black \& Dalgarno (\cite{blackdal}). 
 We use the method of Wright \& Morton (\cite{wrightmort}) to gain information about the strength $U_{\lambda}$ of the FUV radiation field at $\lambda\approx1000$~{\AA}.
 The validity of this consideration depends on the assumption that the bulk of H$_2$ in $J=0$ and $J=2$ is located in the same region as the bulk of H$_2$ in levels $J\geq4$.
 In principle the presence of two clouds with different gas temperatures  could result in a level population as shown here, but this would then be so on all galactic (Savage et al. \cite{savage77}) and Magellanic Cloud (Richter \cite{richter2000} lines of sight investigated.    
 We use the equilibrium equation:
\begin{equation}
p_4\cdot\sum_{J=0,2,4}{(U_{\lambda}W_{\lambda}c/h{\nu})} = {N_{J=4}}{A_4}+\sum_{J=4}{(U_{\lambda}W_{\lambda}c/h{\nu})}
\end{equation}
 The term on the left side of the equation is the number of photo-excited molecules cascading with probability $p_4$ into $J=4$, that on the right side describes the depopulation of the $J=4$ level by radiative decay (probability $A_4$) and photo-excitation. 
 The $W_{\lambda}$ values are calculated with the column densities and the optical depth relation derived from the absorption spectrum, implying that the exciting radiation field comes mainly from \object{Sk\,-69\,246} itself or at least from the same direction.
 Setting $U_{\lambda}=const.$ and solving the equation yields $U_{\lambda}\simeq3\cdot10^{-15}$~erg\,s$^{-1}$\,cm$^{-3}$ (at $\lambda\approx1000$~{\AA}), that is about $8$ times more than derived by Wright \& Morton (\cite{wrightmort}) for the molecular gas in front of $\zeta$ Ophiuchi and $40$ times the typical radiation field found on sight-lines in the galactic disk.  
 Uncertainties in this calculation arise from strong $J=1$ lines reducing the number of photons which can be absorbed by molecules with $J=0$ and absorption by dust within the cloud.
 The consequence is an underestimate of the radiation field at the border of the cloud.

 The derived radiation field can be used for an estimate of the density in Cloud C.
 UV-pumping and collisional processes compete in the population of the $J=2$ level which is raised slightly above the population expected for the kinetic temperature of about $72$~K.
 
 The equilibrium equation for population and depopulation of the $J=2$ level is\begin{eqnarray}
p_2\cdot\sum_{J=0,2}{U_{\lambda}{\cdot}W_{\lambda}{\cdot}c/(h{\cdot}\nu)}+\sigma_{02}{\cdot}n_{\rm H}{\cdot}N_{J=1}= \\ 
\sum_{J=2}{U_{\lambda}{\cdot}W_{\lambda}{\cdot}c/(h{\cdot}\nu)}+\sigma_{20}{\cdot}n_{\rm H}{\cdot}N_{J=2}+N_{J=2}{\cdot}A_{2}\nonumber 
\end{eqnarray}  
 The difference to Eqn. 1 are the terms for collisional excitation/deexcitation of H$_2$ in $J=0$ and $J=2$ by \ion{H}{i} with rate coefficients $\sigma_{02}$ and $\sigma_{20}$ and the number density $n_{\rm H}$ of neutral hydrogen in the region containing H$_2$. 
 Of course a more detailed model would require the consideration of H$_2$, proton, and helium number densities, but for a rough estimate the equation above should be sufficient.
 Using an $A_{2}$ value from Wagenblast (\cite{wagenblast}) and rate coefficients from Sun \& Dalgarno (\cite{sundalg}) we derive values for n$_{\rm H}$ between $5$ and $280$~cm$^{-3}$, assuming a gas temperature of $70$ - $80$~K. 
 
 Under the assumption that the radiation field in cloud C is dominated by \object{Sk\,-69\,246} it is also possible to estimate the distance between the cloud and the star.
 We use $I_{\rm observed}/I_{\rm cloud} = (d/r_{\rm LMC})^2$ with $I_{\rm observed}$ being the intensity at 1000~{\AA} observed in the {\sc Fuse} spectrum, $I_{\rm cloud}$ the intensity in cloud C as derived from the H$_2$ excitation, $r_{\rm LMC}$ the distance of the \object{LMC} (50~kpc), and $d$ the distance between cloud C and \object{Sk\,-69\,246}.
 $I_{\rm observed}$ is corrected for extinction by applying the model by Cardelli et al. (\cite{cardelli}) with $A_V=3.1$.  
 The result depends on the assumed extinction, the larger $E(B-V)$ the larger is the derived distance.  
 We obtain $d\approx15\pm5$~pc for $E(B-V)=0.10_{-0.02}^{+0.05}$.
 Optical images of the \object{30~Dor} region show an \ion{H}{ii} region of roughly $30$ pc radius around \object{Sk\,-69\,246}.
 If a spherical symmetry of the \ion{H}{ii} region around \object{Sk\,-69\,246} is assumed, the extent of this region along the line of sight and our distance estimate are at least of the same order of magnitude. 
 So cloud C might indeed be located rather close to that \ion{H}{ii} region, despite its low temperature and molecular contents.
 The lack of molecular hydrogen, the non-detection of \ion{C}{i}, and the presence of higher ionized gas places cloud B even closer to the star,
 a conclusion already drawn by de Boer et al. (\cite{deboer80}) from an analysis of {\sc IUE} spectra.
 
 The presence of CO in such an inhospitable environment seems a bit surprising. Even more as its abundance relative to H$_2$, $\log$($N$(CO)/$N$(H$_2$))$=-6.6$, is in good agreement with that measured for galactic sight-lines with similar H$_2$ column densities. For an analysis of {\sl Copernicus} observations of galactic CO see Federman et al. (\cite{federman}).
 Savage et al. (\cite{savage77}) investigated H$_2$ towards galactic stars with {\sl Copernicus}. 
 They discovered that at $E(B-V)\simeq0.08$ a transition from low to high H$_2$ column densities occurs.
 This transition was also found to be present in \object{LMC} gas (Richter \cite{richter2000}).
 If the H$_2$ column density depends merely on the total dust amount in a cloud, the H$_2$ - $E(B-V)$ correlation in a metal poor environment like the \object{LMC} should be similar to the galactic one.
 The colour excess within the \object{LMC} towards \object{Sk\,-69\,246} is rather uncertain since neither the total $E(B-V)$ (literature values ranging from 0.08 to 0.15, see Crowther \& Smith \cite{crowther} for references) nor the galatic fraction of $E(B-V)$ (about 0.06, scattering between 0.0 and 0.15, see Oestreicher et al. \cite{oestreicher}) seem to be reliable.
 Applying the average \object{LMC} gas-to-dust ratio $N$(\ion{H}{i})/$E(B-V)=2\cdot10^{22}$~cm$^{-1}$\,mag$^{-1}$ (Koornneef \cite{koornneef}) yields $E(B-V)\approx0.045$ for component C.
 For such a low $E(B-V)$, $\log N$(H$_2)=19.63$ would be extraordinarily high.

 The fractional abundance of H$_2$, calculated with $N({\rm H})=8\cdot10^{20}$~cm$^{-2}$ derived from the cloud C \ion{S}{ii} column density, is $f=2N({\rm H}_2)/[N({\rm \ion{H}{i}})+2N({\rm H}_2)]\approx0.1$, related to the total \object{LMC} H column density it is $f\approx0.07$.
 In a {\sc Fuse} survey of molecular hydrogen in the Magellanic Clouds (Tumlinson et al. \cite{tumlinson}) a few other \object{LMC} sight lines with comparable fractional abundances but higher $E(B-V)$ were found.

\section{Final discussion}
 That the CO to H$_2$ ratio in this \object{LMC} environment has a galactic value is as surprising as the high fractional abundance of H$_2$, considering the strong radiation field and the lower metallicity.
 A reduction of both quantities would be expected from qualitative considerations.
 Lower metallicity means less dust, resulting in a lower H$_2$ column density because the dust surface is needed for its formation. 
 In diffuse clouds H$_2$ self shielding is more important than shielding by dust.
 The CO column density is influenced by the lower metallicity and less H$_2$ in several ways:
the lower abundance of C and O means less CO formation, the gas phase formation of CO depends on $n^2$(H$_2$) (see Federman et al. \cite{federman}) and the shielding of CO  by H$_2$ and dust against dissociative radiation is less efficient. 
 Strangely enough, on this line of sight both H$_2$ and CO seem not to be depleted at all.

 Possibly the abundances of CO and H$_2$ in this cloud are not in equilibrium.
 The cloud might have been larger in the past and now be in a process of disruption by the radiation and a shock caused by the local \ion{H}{ii} region.

\acknowledgements
We thank O. Marggraf for many interesting discussions.
HB is supported by the GK {\it The Magellanic Clouds and other dwarf galaxies}.
Based on observations made with the NASA-CNES-CSA Far Ultraviolet 
Spectroscopic Explorer. FUSE is operated for NASA by the Johns Hopkins 
University under NASA contract NAS5-32985. 
We also made use of the public archive of IUE spectra.

\end{document}